\documentclass[aps,pra,twocolumn,showpacs,superscriptaddress,showkeys,10pt]{revtex4-1}

\usepackage{graphicx}
\usepackage{amsmath}
\usepackage{amssymb}
\usepackage{bm} 
\usepackage{color}

\newcommand{\be}{\begin{equation}}
\newcommand{\ee}{\end{equation}}
\newcommand{\ba}{\begin{eqnarray}}
\newcommand{\ea}{\end{eqnarray}}

\newcommand{\br}{{\mathbf{r}}}
\newcommand{\vect}[1]{\mathbf{#1}}

\newcommand{\spinor}{\vec{\Psi}}
\newcommand{\fmatrix}{\mathbf{F}}
\newcommand{\ev}[1]{\langle{#1}\rangle}
\newcommand{\bias}{B^\mathrm{b}_z}

\begin{document}

\setlength{\arraycolsep}{1.5pt}

\title{Vortex pump for Bose--Einstein condensates utilizing a time-averaged orbiting potential trap}
\date{\today}
\author{Pekko Kuopanportti}\email{pekko.kuopanportti@gmail.com}
\affiliation{QCD Labs, COMP Centre of Excellence, Department of Applied Physics, Aalto University, P.O. Box 13500, FI-00076 Aalto, Finland}
\author{Brian P. Anderson}
\affiliation{College of Optical Sciences, University of Arizona, Tucson, Arizona 85721, USA}
\author{Mikko M\"ott\"onen}
\affiliation{QCD Labs, COMP Centre of Excellence, Department of Applied Physics, Aalto University, P.O. Box 13500, FI-00076 Aalto, Finland}
\affiliation{Low Temperature Laboratory~(OVLL), Aalto University, P.O. Box 13500, FI-00076 Aalto, Finland}

\begin{abstract}
We show that topological vortex pumping can be implemented for a dilute Bose--Einstein condensate confined in a magnetic time-averaged orbiting potential trap with axial optical confinement. Contrary to earlier proposals for the vortex pump, we do not employ an additional optical potential to trap the condensate in the radial direction, but instead, the radial confinement is provided by the magnetic field throughout the pumping cycle. By performing numerical simulations based on the spin-1 Gross--Pitaevskii equation, we find that several pumping cycles can be carried out to produce a highly charged vortex before a majority of the particles escape from the trap or before the vortex splits into singly charged vortices. On the other hand, we observe that an additional, relatively weak optical plug potential is efficient in preventing splitting and reducing particle loss. With these results, we hope to bring the vortex pump closer to experimental realization. 
\end{abstract}

\maketitle

\section{Introduction}\label{sc:intro}

Bose--Einstein condensation in alkali-metal gases was observed experimentally in 1995~\cite{And1995.Sci269.198,Bra1995.PRL75.1687,Dav1995.PRL75.3969,Bra1997.PRL79.1170}. A few years later, these pioneering experiments were followed by the creation of singly quantized vortices~\cite{Mat1999.PRL83.2498,Mad2000.PRL84.806} and vortex lattices~\cite{Mad2000.JMO47.2715,Abo2001.Sci292.476,Ram2001.PRL87.210402} in such systems. Since then, the study of vortices in Bose--Einstein condensates (BECs) has flourished both theoretically and experimentally~\cite{Fet2009.RMP81.647,And2010.JLTP161.574} due to their close connection with phase coherence and superfluidity. In particular, their stability has been the subject of extensive research~\cite{Dod1997.PRA56.587,Rok1997.PRL79.2164,Pu1999.PRA59.1533,Iso1999.PRA60.3313,Svi2000.PRL84.5919,Vir2001.PRL86.2704,Sim2002.PRA65.033614,Kaw2004.PRA70.043610,Jac2005.PRA72.053617,Huh2006.PRA74.063619,
Lun2006.PRA74.063620,Cap2009.JPB42.145301,Kuo2010.PRA81.023603}. 

In a loop encircling a quantized vortex, the phase of the condensate order parameter undergoes an integer multiple $\kappa$ of $2\pi$ windings. In principle, a vortex in a BEC can have any winding number $\kappa$. However, it is well known that a vortex with $|\kappa|>1$ typically has a higher energy than the corresponding number of separated single-quantum vortices. Consequently, vortices with large winding numbers are prone to splitting~\cite{Shi2004.PRL93.160406,Mot2003.PRA68.023611,Gaw2006.JPhysB39.L225,Huh2006.PRL97.110406,Mat2006.PRL97.180409,Iso2007.PRL99.200403,Kar2009.JPhysB42.095301,Kuo2010.PRA81.033627}, which renders them challenging to create with dynamical methods, such as using a focused laser beam to stir~\cite{Mad2000.PRL84.806} or slice through~\cite{Ino2001.PRL87.080402} the BEC, rotating it with an asymmetric trap potential~\cite{Hod2001.PRL88.010405}, or colliding condensates separated by tailored optical potentials~\cite{Sch2007.PRL98.110402}. Being able to produce vortices with large winding numbers would provide access to novel vortex splitting patterns beyond the typical linear chain that prevails for $|\kappa|\leq 4$~\cite{Shi2004.PRL93.160406,Iso2007.PRL99.200403}. Due to the distinct nature of the different splitting patterns predicted for large values of $\kappa$~\cite{Kuo2010.PRA81.033627}, observing the decay of such vortices would allow for a lucid comparison between theory and experiment. Moreover, it has been speculated that giant-vortex splitting may create necessary conditions for the initialization of superfluid turbulence~\cite{Abr1995.PRB52.7018,Ara1996.PRB53.75}.

In addition to the above-mentioned dynamical methods, vortices can be created with the so-called topological phase engineering technique~\cite{Nak2000.PhysicaB284.17,Iso2000.PRA61.063610,Oga2002.PRA66.013617,Mot2002.JPCM14.13481} (see Ref.~\cite{Pie2008.book.vortices} for review), in which the spin degree of freedom of the BEC is controlled adiabatically by a time-dependent nonuniform magnetic field. Since the method does not rely on the relaxation of condensate dynamics, it is especially well suited for producing multiquantum vortices. Indeed, the first two-quantum and four-quantum vortices in dilute BECs were created by applying the technique to spin-1 and spin-2 BECs confined in a magnetic Ioffe--Pritchard (IP) trap~\cite{Lea2002.PRL89.190403}. 

Subsequent theoretical studies have demonstrated that the topological phase engineering technique can also be used to implement a so-called vortex pump~\cite{Mot2007.PRL99.250406,Xu2008.PRA78.043606,Xu2008.NJP11.055019,Xu2010.PRA81.053619,Kuo2010.JLTP161.561}. In this device, a fixed amount of vorticity is added to the BEC in each control cycle, and thus its repeated application would---stability issues notwithstanding---enable the creation of vortices with arbitrarily large winding numbers. The original proposal~\cite{Mot2007.PRL99.250406} involved creating $2F$ quanta of vorticity per cycle in a spin-$F$ BEC with a magnetic-field configuration consisting of the standard IP trap and an additional hexapole magnetic field. Backed by numerical simulations, the pump was shown to be operable both fully adiabatically and partly nonadiabatically. Later, Xu \emph{et al.}~\cite{Xu2008.PRA78.043606} presented a different pumping cycle for the IP trap in which the hexapole field was replaced with a uniform transverse field. Unfortunately, both of these control cycles suffer from the fact that the magnetic fields provide radial confinement only during part of the cycle, and thus, the fully adiabatic operation of the pump necessitates an optical trap to confine the BEC radially. Since the purpose of the IP trap has been to confine the atomic cloud in the first place, there has been little incentive to supplement it with an optical trap or a hexapole field. Therefore, pumping schemes not requiring such extra ingredients would be desirable from a practical standpoint.

Recently, Xu~\emph{et al.} showed theoretically that vortex pumping can be applied to quantum superpositions to generate countercirculation states~\cite{Xu2010.PRA81.053619}. The authors used a novel control cycle which is particularly suitable for the time-averaged orbiting potential (TOP) trap~\cite{Pet1995.PRL74.3352} and has the advantage that the radially confining quadrupole field can be kept on throughout the entire cycle. However, since only one of the components in the superposition state could be trapped magnetically, a three-dimensional optical trap, as well as a strong optical plug potential piercing the vortex core, had to be employed~\cite{Xu2010.PRA81.053619}.

The aim of this article is to bring the vortex pump closer to experimental realization by showing that it can be implemented with mature, existing technologies alternative to the ones considered in Refs.~\cite{Mot2007.PRL99.250406,Xu2008.PRA78.043606,Xu2008.NJP11.055019,Xu2010.PRA81.053619,Kuo2010.JLTP161.561}. To this end, we demonstrate that vortices can be efficiently pumped in the TOP trap without using, in contrast to Ref.~\cite{Xu2010.PRA81.053619}, additional optical potentials to confine the BEC in the radial direction or to pin the vortex core. Instead, the radial confinement is provided solely by the magnetic field throughout the pumping process, and optical trapping is required only in the axial direction. We present simulations based on the Gross--Pitaevskii equation which indicate that several pumping cycles can be carried out before a majority of the particles escape from the trap or before the generated multiquantum vortex splits into singly quantized vortices due to dynamical instabilities~\cite{Kuo2010.PRA81.033627,Pie2007.PRA76.023610}. On the other hand, we also show that even a relatively weak optical plug potential is efficient in preventing the splitting and in reducing the loss of particles, thereby enabling the controlled creation of isolated vortices with large winding numbers.

The remainder of this article is organized as follows. In Sec.~\ref{sc:theory}, we present the zero-temperature mean-field theory of the spin-1 BEC, describe the control cycle of the vortex pump, and discuss in detail the confinement of the condensate during the cycle. Section~\ref{sc:results} presents our numerical results, which we relate to realistic experimental setups in Sec.~\ref{sc:experimental}. Finally, Sec.~\ref{sc:conclusion} concludes the article with a discussion.

\section{Theory and methods}\label{sc:theory}

\subsection{Mean-field model}

We consider a dilute spin-1 BEC in the zero-temperature limit, thereby neglecting the possible effects due to noncondensed atoms. In the standard mean-field treatment, the spin-1 condensate is described by a three-component order-parameter field that we write in the eigenbasis of the spin-1 matrix $F_z$ as $\spinor=\left(\Psi_{+1},\Psi_{0},\Psi_{-1}\right)$. Its time dependence is given by the spin-1 Gross--Pitaevskii (GP) equation~\cite{Ohm1998.JPSJ67.1822,Ho1998.PRL81.742}
\begin{eqnarray}
\label{eq:GPE}
i\hbar\partial_t\spinor(\br,t) &=& \Big( {\mathcal H} + g_\mathrm{n} \spinor^\dagger\spinor \nonumber \\
&&  + g_\mathrm{s} \spinor^\dagger \fmatrix \spinor \cdot \fmatrix \Big) \spinor(\br,t).
\end{eqnarray}
The single-particle Hamiltonian operator ${\mathcal H}$ is given by
\be\label{eq:ham}
{\cal H} = -\frac{\hbar^2}{2m}\nabla^2 + V_\mathrm{opt}(\br) + \mu_\mathrm{B} g_F \vect{B}\left(\br,t\right)\cdot \fmatrix,
\ee
where $m$ denotes the atomic mass, $g_F$ is the Land\'{e} factor, $\mu_\mathrm{B}$ is the Bohr magneton, $\vect{B}(\br,t)$ denotes the external magnetic field, and $\vect{F}=\left(F_x,F_y,F_z\right)$ is a vector of the standard spin-1 matrices~\cite{Ued2010.book.Bose}. Optical potential terms are contained in $V_\mathrm{opt}(\br)= V_\mathrm{tr}(z) + V_\mathrm{plug}\left( r \right)$, where $ V_\mathrm{tr}(z)=m\omega_z^2 z^2/2$ is a strong axial harmonic trap and $V_\mathrm{plug}\left(r\right)=A \exp\left(-r^2/d^2\right)$ describes a possibly present Gaussian-shaped repulsive plug potential of amplitude $A\geq 0$ and width $d$. Here, $r=\sqrt{x^2+y^2}$ is the radial coordinate. The coupling constants $g_\mathrm{n}$ and $g_\mathrm{s}$ appearing in Eq.~\eqref{eq:GPE} measure the strengths of the local density--density and spin--spin interactions, respectively. They are related to the $s$-wave scattering lengths $a^{(0)}$ and $a^{(2)}$ into spin channels with total spin 0 and $2\hbar$ by the expressions $g_\mathrm{n}=4\pi\hbar^2\left[a^{(0)}+2a^{(2)}\right]/3m$ and $g_\mathrm{s}=4\pi\hbar^2\left[a^{(2)}-a^{(0)}\right]/3m$. The order parameter is normalized such that $\int d^3 r \spinor^\dagger \spinor = N_0$, where $N_0$ is the number of particles in the BEC.

\subsection{Magnetic fields and the pumping cycle}

The operation principle of the vortex pump is to control the spin degree of freedom of the condensate locally by slowly tuning the magnetic field $\vect{B}(\br,t)$ in a cyclic manner such that the system acquires a fixed amount of vorticity per cycle~\cite{Mot2007.PRL99.250406,Xu2008.PRA78.043606,Xu2008.NJP11.055019,Xu2010.PRA81.053619,Kuo2010.JLTP161.561}. In the pumping scheme considered here, the spin-1 atoms are assumed to be magnetically confined in the standard TOP trap~\cite{Pet1995.PRL74.3352}. It consists of a quadrupole field $\vect{B}_\mathrm{q}$, which has axial symmetry about the $z$ direction, and a rapidly rotating, spatially uniform magnetic field $\vect{B}_\mathrm{rot}$ oriented along the $xy$ plane. In addition, we assume that the TOP trap is accompanied by a uniform axial bias field $\bias(t) \hat{\vect{z}}$ that can be controlled independently of the other fields. The total magnetic field can be written as
\begin{equation}\label{eq:B}
\vect{B}(\br,t) = \vect{B}_\mathrm{q}(\br) + \vect{B}_\mathrm{rot}(t)+\bias(t)\hat{\vect{z}},
\end{equation}
where $\vect{B}_\mathrm{q}(\br) = B'\left(x\hat{\vect{x}}+y\hat{\vect{y}}-2 z \hat{\vect{z}}\right)$ is the quadrupole field with the radial gradient $B'$ and the rotating transverse bias field is given by
\begin{equation}\label{eq:Brot}
\vect{B}_\mathrm{rot}(t) = B_\mathrm{rot}(t) \left[\cos\left(\omega_\mathrm{rot} t\right) \hat{\vect{x}}+\sin\left(\omega_\mathrm{rot} t\right)\hat{\vect{y}} \right],
\end{equation}
where $\omega_\mathrm{rot}$ denotes its angular frequency of rotation about the $z$ axis. The bias field strengths are assumed to be bound by $B_0$ such that $\bias(t)\in\left[-B_0, B_0\right]$ and $B_\mathrm{rot}(t)\in \left[0, B_0 \right]$. We point out that the field configuration of Eq.~\eqref{eq:B} has already been employed in BEC experiments~\cite{Hod2000.JPhysB33.4087,Hod2001.PRL86.2196}. 

\begin{figure}
\begin{center}
\includegraphics[
  width=210pt,
  keepaspectratio]{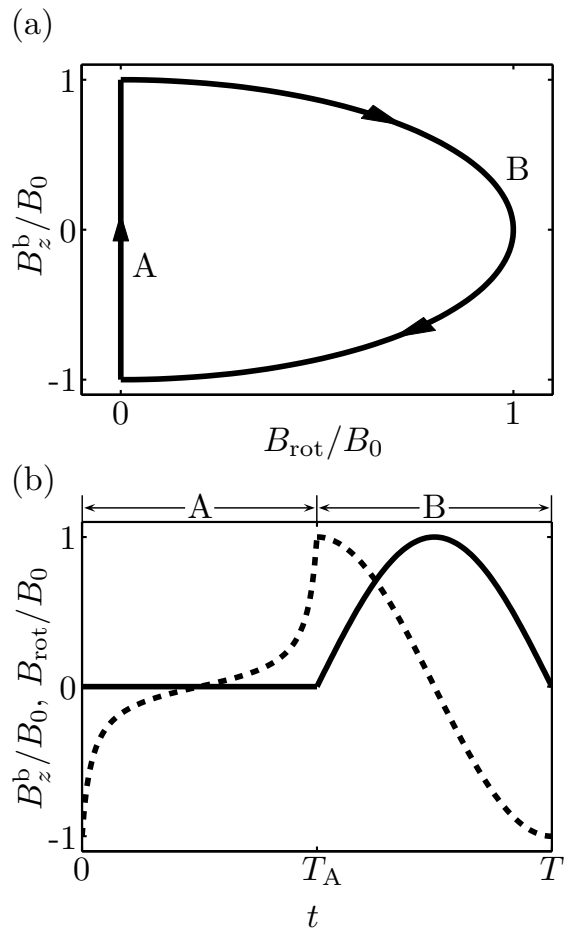}
\end{center}
\caption{\label{fig:cycle} (a) Control cycle of the vortex pump in the $(B_\mathrm{rot},\bias)$ plane, 
where $B_\mathrm{rot}$ and $\bias$ are the strengths of the transverse and axial magnetic bias fields, respectively. 
The cycle starts at $(B_\mathrm{rot},\bias)=(0,-B_0)$ and proceeds clockwise. (b) Time dependence of $B_\mathrm{rot}$ (solid line) and $\bias$ (dashed line) [Eqs.~(\ref{eq:part_A}) and (\ref{eq:part_B})] during the cycle. 
The cycle is divided into parts A and B as indicated. 
In part A, the axial bias field $\bias$ can also be reversed linearly. }
\end{figure}

In order to facilitate vortex pumping in the TOP trap, we use the control cycle presented in Fig.~\ref{fig:cycle}. It is carried out by tuning two magnetic-field parameters, $\bias$ and $B_\mathrm{rot}$, and can be divided into part A ($0\leq t \leq T_\mathrm{A}$) and part B ($T_\mathrm{A} \leq t \leq T_\mathrm{A}+T_\mathrm{B} =: T$). Part A is similar to the original proposals of topological phase engineering~\cite{Nak2000.PhysicaB284.17,Iso2000.PRA61.063610,Oga2002.PRA66.013617,Mot2002.JPCM14.13481,Mot2007.PRL99.250406} and the experiments~\cite{Shi2004.PRL93.160406,Lea2002.PRL89.190403,Lea2003.PRL90.140403,Kum2006.PRA73.063605,Oka2007.JLTP148.447,Shi2011.JPB44.075302}, and it is responsible for increasing the circulation in the spin-1 BEC by two quanta. It is executed by reversing the axial bias field with the rotating field switched off, 
\be\label{eq:part_A}
\left. \begin{array}{ll} B_\mathrm{rot}(t) = 0 \\ \bias(t)= B' \rho_0 \tan\left[\frac{2t-T_\mathrm{A}}{T_\mathrm{A}}\arctan\left(\frac{B_0}{B'\rho_0} \right) \right] \end{array}\right\}\ 0\leq t \leq T_\mathrm{A},
\ee
where $B_0$ should be large enough to render the BEC essentially spin polarized along the $z$ axis at $t=0$ and $t=T_\mathrm{A}$. To improve adiabaticity, the time dependence for $\bias$ has been chosen such that spins at a distance of $\rho_0$ from the $z$ axis are turned with constant speed, but part A can also be performed by reversing $\bias(t)$ linearly in time~\cite{Lea2002.PRL89.190403}. In part B, the axial bias field is returned to its initial value while ramping up and down the rotating field, 
\be\label{eq:part_B}
\left. \begin{array}{ll} B_\mathrm{rot}(t) = B_0\sin\beta(t) \\ \bias(t)=B_0 \cos\beta(t) \end{array}\right\}\ T_\mathrm{A} \leq t \leq T,
\ee
where $\beta(t)=\pi\left(t-T_\mathrm{A}\right)/T_\mathrm{B}$. Part B was originally proposed by Xu~\emph{et al.}~\cite{Xu2010.PRA81.053619}, and it is designed to preserve the accumulated vorticity. The cyclic repetition of parts A and B will therefore increase the vortex winding number of the spin-1 BEC by two per cycle.

To efficiently steer the condensate spin by the magnetic field $\vect{B}\left(\br,t \right)$ requires that the Zeeman energy dominates over the kinetic energy at each point in space. Hence, to guarantee adiabaticity,  $|\vect{B}\left(\br,t \right)|$ should be sufficiently large in the region occupied by the BEC. This condition is not fulfilled at the origin when $\bias$ crosses zero at $t=T_\mathrm{A}/2$, and thus it is desirable to prevent particles from entering this area. This can be accomplished by introducing the repulsive plug potential $V_\mathrm{plug}(r)$ along the $z$ axis. The plug not only improves adiabaticity but also serves to stabilize the created multiquantum vortex against splitting~\cite{Kuo2010.PRA81.033627,Kuo2010.JLTP161.561}. The plug can be realized for pancake-shaped BECs by a focused blue-detuned laser beam as has been done in various experiments~\cite{Dav1995.PRL75.3969,Abo2001.Sci292.476,Ram2001.PRL87.210402,Sim2005.PRL94.080404,Nee2010.PRL104.160401}. 
In this article, we present results for vortex pumping both with and without the plug potential. 

\subsection{Confinement during pumping}\label{subsc:confinement}

An essential difference between this article and earlier work concerning the vortex pump~\cite{Mot2007.PRL99.250406,Xu2008.PRA78.043606,Xu2008.NJP11.055019,Xu2010.PRA81.053619,Kuo2010.JLTP161.561} is that here we never employ an optical trapping potential in the radial direction. Instead, radial confinement is provided by the magnetic field throughout the entire pumping cycle. In the case of spin-1 BECs, the magnetically trapped weak-field seeking state (WFSS) corresponds locally to the highest-energy eigenstate of the Zeeman Hamiltonian $g_F \mu_\mathrm{B}\left(\br,t\right) \vect{B} \cdot \fmatrix$, with the effective trap potential given by the local eigenvalue  $|g_F\mu_\mathrm{B}\vect{B}(\br,t)|$. Even though this potential is solely responsible for the radial confinement, a strong optical trap is still needed in the axial direction to keep the atomic cloud centered around $z=0$ throughout the pumping cycle. Changes in $\bias$ shift the $z$ coordinate of the zero-value point of the total magnetic field, and without the optical $z$ confinement, carrying out the cycle would merely move the whole BEC along the $z$ axis. Hence, the Hamiltonian in Eq.~\eqref{eq:ham} includes the harmonic axial trapping potential $V_\mathrm{tr}(z) = m\omega_z^2 z ^2 / 2$ with a trap frequency $\omega_z$ that is assumed to be large enough to render the condensate pancake shaped, which means that the order parameter can be taken to have the form $\spinor\left(\br,t\right)=\spinor_\mathrm{2D}\left(x,y,t\right)\zeta(z)$, where $\zeta(z)=\exp\left(-z^2/2 a_z^2\right)/\sqrt[4]{\pi a_z^2}$ and $a_z = \sqrt{\hbar/m\omega_z}$ is the axial oscillator length. This enables us to integrate out the $z$ variable in Eq.~\eqref{eq:GPE} and obtain an effectively two-dimensional GP equation with the magnetic field determined at $z=0$.

Let us consider the shape of the magnetic potential in the vicinity of the origin. During part A of the cycle ($0\leq t \leq T_\mathrm{A}$), the strength of the magnetic field is given by
\begin{eqnarray}
|\vect{B}| &=&\sqrt{(B'x)^2+(B'y)^2+\left[\bias-2B'z\right]^2}\nonumber  \\  &\approx& |\bias|- \frac{2B'|\bias|}{\bias}z +\frac{B'^2}{2|\bias|} r^2,
\label{eq:BstrengthA}\end{eqnarray}
where in the expansion we have neglected third- and higher-order terms in $B'r/|\bias|$ and $B'|z|/|\bias|$. Therefore, the magnetic field at $t=0$ gives rise to an approximately harmonic potential in the radial direction with the trap frequency
\be\label{eq:omega0}
\omega_0 = B'(0) \sqrt{\left|\frac{g_F \mu_\mathrm{B}}{m \bias(0)}\right|}.
\ee 
It is convenient to measure all quantities in terms of $\omega_0$, and thus we express lengths in units of the corresponding oscillator length $a_0=\sqrt{\hbar/m \omega_0} \gg a_z$, energies in units of $\hbar\omega_0$, time in units of $1/\omega_0$, and the magnetic field in units of $\hbar\omega_0/|g_F|\mu_\mathrm{B}$. Variables expressed in these units are henceforth denoted with a tilde.

According to Eq.~\eqref{eq:BstrengthA}, the profile of the radial confinement will change during part A. Initially, the potential is harmonic, with the effective trap frequency $\propto |\bias(t)|^{-1/2}$ increasing in time. At $t=T_\mathrm{A}/2$, the trap becomes purely linear in $r$ with the gradient $|g_F|\mu_\mathrm{B} B'$. The axisymmetric modulations of the trap profile will cause shrinking of the BEC and excitation of its breathing mode. Although these effects do not critically hinder the operation of the pump, they can nevertheless be reduced by introducing time dependence into $B'$ such that it is decreased when $|\bias|$ is ramped down during part A. Here, we use the dependence
\begin{equation}\label{eq:B'}
B'(t)=\left\{\begin{array}{ll} B_0'\left[\frac{B_\mathrm{min}'}{B_0'}+\left(\frac{2t-T_\mathrm{A}}{T_\mathrm{A}}\right)^2\left(1-\frac{B'_\mathrm{min}}{B_0'}\right)\right], & 0 \leq t \leq T_\mathrm{A}, \\ B'_0, & T_\mathrm{A} < t \leq T. \end{array} \right.
\end{equation}
with $B_0'=B'(0)$ and $B_\mathrm{min}' \approx 0.4 B'_0$.  Varying $B'$ is not necessary but it improves the accuracy of the pump and reduces the loss of particles. 

In part B, the frequency $\omega_\mathrm{rot}$ of the rotating bias field is chosen to be low compared with the frequencies of transitions between different magnetic substates but large compared with the effective radial trap frequency. Typically, magnetic trap frequencies are of order $10^2\,\mathrm{Hz}$ while the transition frequencies are of order $10^6\,\mathrm{Hz}$. Thus, a reasonable choice would be, e.g., $\omega_\mathrm{rot} \sim 10^4\,\mathrm{Hz}$. These conditions ensure that the atoms will not undergo transitions to other substates and be lost from the magnetic trap but instead move in an effective potential given by the time average of the instantaneous magnetic potential over one rotation period of the field $\vect{B}_\mathrm{rot}$. For $T_\mathrm{A} \leq t \leq T$, the fast-time-averaged field strength can be expanded in a power series in $B'r/B_0$ and $B'|z|/B_0$ as
\begin{eqnarray}\label{eq:BstrengthB}
\frac{\omega_{\mathrm{rot}}}{2\pi}\int_{I_t} dt'|\vect{B}(t')| & \approx & B_0 - 2B'z \cos\beta + \frac{B'^2}{4B_0}\left(1+\cos^2\beta\right)r^2 \nonumber \\ &+&
\frac{2 B'^2\sin^2\beta}{B_0} z^2, 
\end{eqnarray}
where the third- and higher-order terms have been discarded and the interval of integration is $I_t=\left[t-\pi/\omega_\mathrm{rot}, t+\pi/\omega_\mathrm{rot}\right]$. Equation~\eqref{eq:BstrengthB} implies that the effective radial trap frequency will decrease by $\sim 29\%$ during part B, but this should not significantly disturb the pumping process.

\section{Results}\label{sc:results}

We study the temporal evolution of a spin-1 BEC during vortex pumping by numerically solving the GP equation, Eq.~\eqref{eq:GPE}, with the $T$-periodic time dependence of the magnetic field $\vect{B}(\br,t)$ given by Eqs.~\eqref{eq:part_A}, \eqref{eq:part_B}, and \eqref{eq:B'}. After factoring out the $z$ dependence of the order parameter as $\spinor\left(\br,t\right)=\spinor_\mathrm{2D}\left(x,y,t\right)\zeta(z)$, Eq.~\eqref{eq:GPE} is discretized on a uniform grid with a finite-difference method and integrated in time for several consecutive pumping cycles using a split-operator approach. The dimensionless coupling constants are chosen to have the values $\tilde{g}_\mathrm{n}= N_0 m   g_\mathrm{n}/\sqrt{2\pi \hbar^4 a_z^2 } =250$ and $\tilde{g}_\mathrm{s}= N_0 m   g_\mathrm{s}/\sqrt{2\pi \hbar^4 a_z^2 } = -0.01\,\tilde{g}_\mathrm{n}$, the latter corresponding to spin-1 condensates of ${}^{87}$Rb~\cite{Kla2001.PRA64.053602,Kem2002.PRL88.093201,Wid2006.NJP8.152,Note1}. The durations for parts A and B of the control cycle are given by $\tilde{T}_A = \omega_0 T_\mathrm{A} = 3$ and $\tilde{T}_\mathrm{B}= \omega_0 T_\mathrm{B} = 2$, respectively, and the parameters in Eqs.~\eqref{eq:part_A}, \eqref{eq:part_B}, and \eqref{eq:B'} have the values $\tilde{B}_0= |g_F|\mu_\mathrm{B}B_0/\hbar\omega_0=200$, $\tilde{\rho}_0=\rho_0/a_0=5$, $\tilde{B}_0'=|g_F|\mu_\mathrm{B} a_0 B'_0/\hbar\omega_0 = \sqrt{200}$, and $\tilde{B}_\mathrm{min}'=|g_F|\mu_\mathrm{B} a_0 B'_\mathrm{min}/\hbar\omega_0  =6$. The Land\'{e} factor $g_F$ is taken to be negative as in the case of spin-1 ${}^{87}$Rb. The frequency of the rotating field is set to $\tilde{\omega}_\mathrm{rot} =\omega_\mathrm{rot}/\omega_0 =85$. We present results both with and without an optical plug potential of amplitude $\tilde{A}=A/\hbar\omega_0 = 10$ and width $\tilde{d}=d/a_0=2$. Before the pumping is started, a relaxation method is used to bring the BEC to the lowest-energy WFSS with the magnetic field in its $t=0$ configuration. 

\begin{figure*}
\begin{center}
\includegraphics[
  width=0.9\textwidth,
  keepaspectratio]{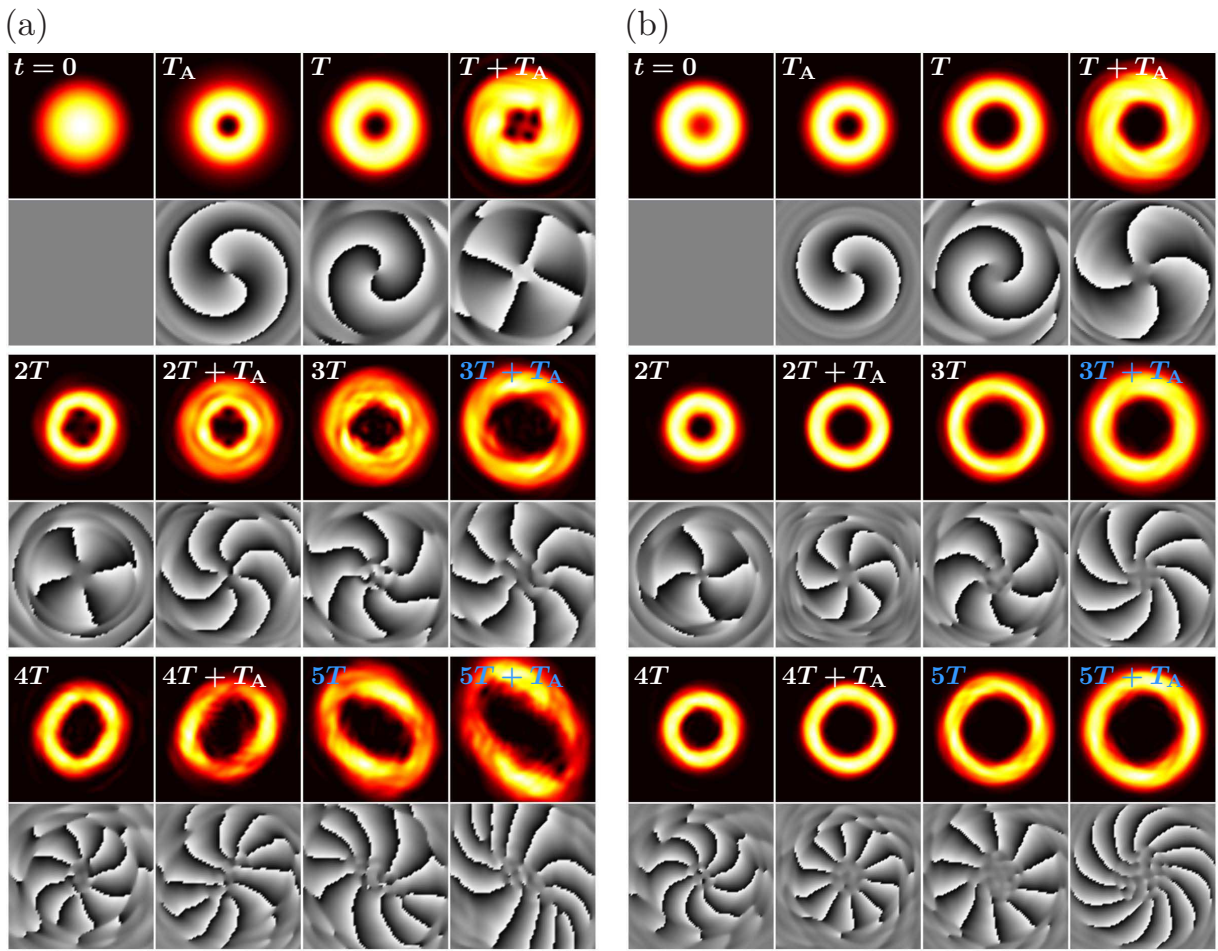}
\end{center}
\caption{\label{fig:T5-profiles}  (Color online) Areal particle density and complex phase of the order-parameter components $\Psi_{+1}$ (at $t=l T$, $l\in\mathbb{N}$) and $\Psi_{-1}$ (at $t=l T + T_\mathrm{A}$) in the $xy$ plane during pumping (a) without and (b) with an optical plug potential of amplitude $A=10\hbar \omega_0$ and width $d=2 a_0$. The field of view in the panels is (a) $12a_0 \times 12a_0$ and (b) $14a_0\times 14 a_0$. The two parts of the control cycle have the durations $T_\mathrm{A}=3/\omega_0$ and $T_\mathrm{B}=2/\omega_0$, and the dimensionless coupling constants are set to $\tilde{g}_\mathrm{n}=250$ and  $\tilde{g}_\mathrm{s}=-0.01 \tilde{g}_\mathrm{n}$.}
\end{figure*}

Figure~\ref{fig:T5-profiles} shows the squared moduli and the complex phases of the most relevant order-parameter components $\Psi_{+1}$ (at $t=l T$, $l\in\mathbb{N}$) and $\Psi_{-1}$ (at $t=l T + T_\mathrm{A}$) during the pumping process.
The accumulation of two quanta of vorticity during part A of each cycle is clearly visible in the phase fields at $t=l T+T_\mathrm{A}$. Part B of the cycle, during which the rotating bias field is on, is observed to leave the vorticity unaffected. The pumping also causes breathing of the BEC, as indicated by its oscillating spatial extent and by the nonzero radial derivatives of the phase fields. The excitation of the breathing mode is attributed to the changing magnetic confinement during the pumping cycle [see Eqs.~\eqref{eq:BstrengthA} and \eqref{eq:BstrengthB}]. 

\begin{figure*}
\begin{center}
\includegraphics[
  width=1.0\textwidth,
  keepaspectratio]{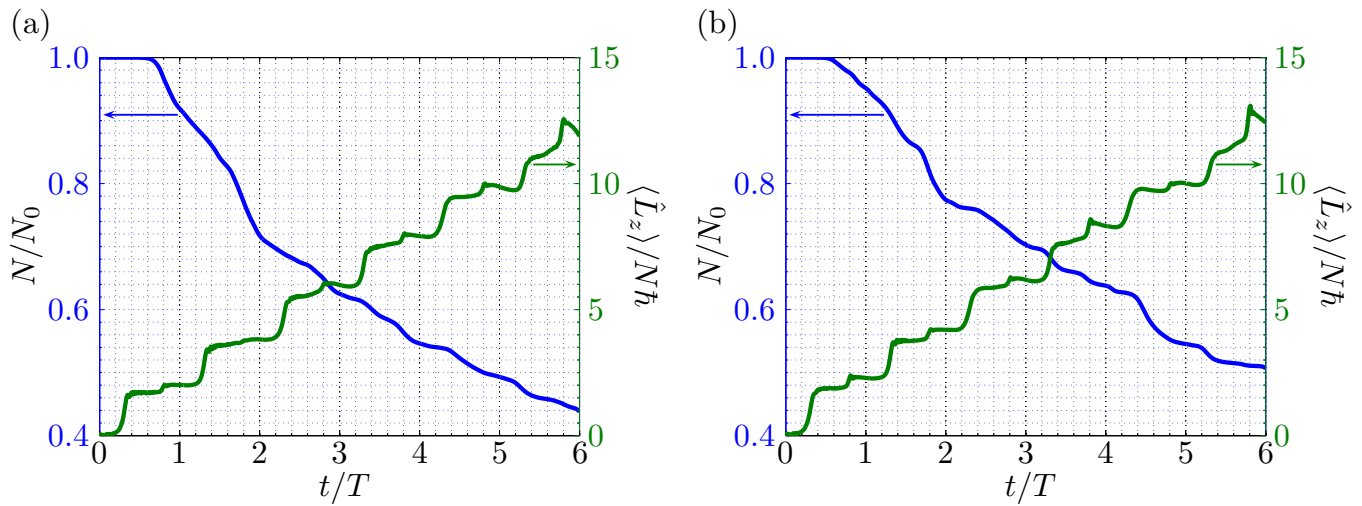}
\end{center}
\caption{\label{fig:T5-L-N} (Color online) Particle number $N$ and average orbital angular momentum $\langle \hat{L}_z \rangle/N$ of the BEC as functions of time during the vortex pump simulations presented in Fig.~\ref{fig:T5-profiles}: (a) no optical plug potential; (b) Gaussian-shaped plug of amplitude $A=10\hbar\omega_0$ and width $d=2 a_r$. }
\end{figure*}

Axisymmetric vortex states with large winding numbers $\kappa$ have been found to be dynamically unstable against splitting in pancake-shaped, harmonically trapped single-component BECs for most values of the interatomic interaction strength, with the degree of instability generally increasing with increasing $\kappa$~\cite{Kuo2010.PRA81.023603,Kuo2010.PRA81.033627}. Therefore, when the stabilizing plug potential is not employed in the pumping, the created multiquantum vortex is expected to split after it has accumulated a sufficiently large winding number. In Fig.~\ref{fig:T5-profiles}(a), the onset of splitting is visible around $t=T+T_\mathrm{A}$, when $\kappa=4$. As shown for $t=5T+T_\mathrm{A}$, the process eventually results in a line of singly quantized vortices and is thus ascribed to a dynamically unstable excitation mode with orbital angular momentum of $\pm 2\hbar$ per particle with respect to the condensate~\cite{Kuo2010.PRA81.033627}. On the other hand, when the relatively weak plug is employed [Fig.~\ref{fig:T5-L-N}(b)], the vortex does not split despite its significant breathing, and a nearly symmetric 12-quantum-vortex state is observed at $t=5T+T_\mathrm{A}$. We have confirmed numerically that the plug amplitude $A$ can be subsequently ramped down without destroying the state.

Due to the finite pumping period $T$ and magnetic field strength $|\vect{B}|$, the pumping process is not perfectly adiabatic, and there are spins that do not follow the local magnetic field. Since these spin components are no longer trapped, they escape the condensate region. Consequently, the number of particles in the trap decreases during the process. The loss rate depends on the degree of adiabaticity of the pump, i.e., on the pumping speed and on the local field strength $|\vect{B}(\br)|$. 

Figure~\ref{fig:T5-L-N} presents the number of particles in the trap, $N=\int_{r\leq R} d^3 r \spinor^\dagger\spinor \leq N_0 $, and their average orbital angular momentum $\ev{\hat{L}_z}/N=-i\hbar\int_{r\leq R} d^3 r \spinor^\dagger \left[ \hat{\vect{z}} \cdot \left( \vect{r} \times \nabla \right) \right] \spinor/N$ as functions of time during the pumping process. Here, the particles are considered lost after their distance from the $z$ axis exceeds $R=13 a_0$. Therefore, the curves in Fig.~\ref{fig:T5-L-N} also include contributions from unconfined atoms that have not drifted away from the trap region. Even after taking this into account, we find that a substantial portion of the atoms still remain in the WFSS at $t=5T$. As expected, the optical plug [Fig.~\ref{fig:T5-L-N}(b)] is observed to significantly reduce the loss of particles.

Whereas the number of vortices in a scalar condensate is quantized, the orbital angular momentum $\ev{\hat{L}_z}$ is a continuous quantity. The angular-momentum curves in Fig.~\ref{fig:T5-L-N} indicate the increment of vorticity by two during part A of each cycle and show that $\ev{\hat{L}_z}/N$ increases monotonously during the first few cycles. The considerable deviation of $\ev{\hat{L}_z}$ from the ideal value $2l\hbar N$ after $l$ cycles is mainly due to the contribution of the untrapped atoms that remain in the region $r\leq R$. Moreover, the slight increase in $\ev{\hat{L}_z}/N$ during each part B is attributed to the small center-of-mass motion induced by the rotating transverse bias field.

\section{Experimental feasibility}\label{sc:experimental}

Let us briefly relate the proposed topological vortex pump to a realistic experimental setup. As an example of previously realized experimental parameters, the experiment of Ref.~\cite{Nee2010.PRL104.160401} studied condensates of $2\times 10^6$ $^{87}$Rb atoms in the $5\,^2S_{1/2}\;|F=1,\,m_F=-1\rangle$ state. The atoms were confined in a TOP trap with a bias field of $B_{\mathrm{rot}}=5$~G and a quadrupole field with the radial gradient $B'=27$~G/cm.  An additional red-detuned Gaussian laser beam propagating in the $xy$ plane provided strong confinement along the $z$~axis and negligible confinement in the radial plane; the TOP trap provided the significant portion of radial confinement.  This laser beam had a wavelength of 1090~nm, a power of $\sim$\,0.5~W, and radii of $\sim$\,20~$\mu$m along $z$ and $\sim$\,2~mm along $r$.  The combined optical and magnetic trap gave trapping frequencies of $\left(\omega_0,\omega_z \right) = 2\pi\times \left(8,90\right)$~Hz, low compared with the TOP trap rotation frequency of $\omega_{\mathrm{rot}}=2\pi\times 4$~kHz.

As a proposed implementation of the vortex pump, we consider the spin-1 condensates of ${}^{87}$Rb and the following field parameters.  First, in place of a red-detuned trapping laser, we assume the use of a blue-detuned beam that has a Gaussian profile in the $xy$ plane and a first-order Hermite--Gauss profile along $z$. Since the atoms would be trapped in the dark region between the two halves of the beam, the laser would not provide any radial confinement. A 1-W, 532-nm beam, with Gaussian radii of $\sim$\,9~$\mu$m along $z$ and $\sim$\,500~$\mu$m along $r$, will give an axial trapping frequency of $\omega_z \approx 2\pi \times 900$~Hz. This field remains constant throughout the pumping cycle and is large enough to support the atoms against gravity and the magnetic-field forces along $z$. 

Secondly, for the magnetic-field parameters, we assume the field values given above that correspond to Ref.~\cite{Nee2010.PRL104.160401}. Thus, we take the maximum strength of the uniform fields to be $B_0=5$~G [Eqs.~\eqref{eq:part_A} and \eqref{eq:part_B}], which  is reached for the axial bias field $B_z^\mathrm{b}$ at the beginning and end of part A of the cycle, i.e., at $t=0$ and $t=T_\mathrm{A}$. Between these times, the rotating component of the TOP trap is off, the quadrupole field with the gradient $B'_0=27$~G/cm is on, and the bias field pushes the zero-value point of the total magnetic field $\vect{B}$ from nearly 1~mm above the pancake-shaped BEC to nearly 1~mm below the BEC. For this trap, the harmonic approximation of Eqs.~\eqref{eq:BstrengthA} and~\eqref{eq:BstrengthB} is well justified. Equation~\eqref{eq:omega0} yields $\omega_0 \approx 2\pi \times 11$~Hz, and hence the simulations in Sec.~\ref{sc:results} correspond to $T_\mathrm{A} = 3/\omega_0=44$~ms. As assumed in Eq.~\eqref{eq:B'}, $B'\left(t\right)$ can optionally be ramped to a minimum value of $B'_\mathrm{min} \approx 11$ G/cm during part~A. In part B, the rotating bias field $\vect{B}_\mathrm{rot}$ is ramped on while the bias field $B_z^\mathrm{b}$ reverses direction, bringing the zero-value point of $\vect{B}$ in a spiraling trajectory around the BEC prior to $\vect{B}_\mathrm{rot}$ being ramped back off. Our simulations fix the duration of this stage at $T_\mathrm{B} = 2/\omega_0=29$~ms.

Additionally, for the parameters assumed above, the implementation of an optical plug would be straightforward.  The assumed value of the plug radius used in our simulations, $d=2 a_0$, corresponds to a Gaussian $1/e^2$ beam radius of $2\sqrt{2}a_0 \approx 9$~$\mu$m, similar to the beam used in Ref.~\cite{Nee2010.PRL104.160401}.

For numerical convenience, our simulations have assumed smaller numbers of atoms than would be ideally used in an experiment, as well as lower values of magnetic fields than those typically found in TOP traps~\cite{Note2}. Nevertheless, based on the validity of the harmonic approximation and the readily achievable time and length scales, the primary features seen in the simulations should be preserved and observable with experimentally feasible parameters.

\section{Conclusion}\label{sc:conclusion}

In summary, we have discussed how to implement a vortex pump for a BEC in a TOP trap, resorting only to standard experimental techniques and magnetic-field configurations that are already available in BEC laboratories. We showed that the pumping can be carried out without using an additional optical potential to trap the atoms in the radial direction. Instead, the radial confinement is provided solely by the magnetic field, and a harmonic optical potential is employed only in the axial direction. Our simulations demonstrated that even if the pumped multiquantum vortices are not stabilized by a Gaussian-shaped plug potential piercing their core, several pumping cycles can still be carried out before the vortex splits clearly. On the other hand, already a relatively weak plug potential was found to prevent the splitting and to reduce the loss of atoms from the trap. 

Our results are expected to facilitate the experimental realization of the vortex pump. This achievement would represent an important milestone in vortex physics, since it would provide a controlled method to produce almost any desired amount of vorticity. From a theoretical point of view, the vortex pump is a fascinating example of adiabatic quantum dynamics for which the control parameters of the system are varied cyclically but the system does not return to its initial eigenspace. In fact, the appearance of vortices can be interpreted as the accumulation of a position-dependent geometric Berry phase~\cite{Ber1984.ProcRSocA392.45} for individual spins of the condensate atoms~\cite{Nak2000.PhysicaB284.17,Iso2000.PRA61.063610,Oga2002.PRA66.013617,Mot2002.JPCM14.13481,Lea2002.PRL89.190403,Mot2007.PRL99.250406,Pie2008.book.vortices}.

\begin{acknowledgments}
The authors thank J. A. M. Huhtam\"aki, P.~J. Jones, V.~Pietil\"a, and E.~Ruokokoski for insightful comments and discussions. CSC - IT Center for Science Ltd. is acknowledged for computational resources. P.K. and M.M. have been supported by the Emil Aaltonen Foundation and by the Academy of Finland under Grants No.~135794, 138903, 141015, and through its Centres of Excellence Program under Grant No.~251748 (COMP). B.P.A. acknowledges the support of the US National Science Foundation Grant PHY-1205713. P.K. thanks the Finnish Cultural Foundation, the KAUTE Foundation, and the Magnus Ehrnrooth Foundation for financial support. 
\end{acknowledgments}

\bibliography{toppump}

\begin{thebibliography}{10}

\bibitem{And1995.Sci269.198}
M.~H. Anderson, J.~R. Ensher, M.~R. Matthews, C.~E. Wieman, and E.~A. Cornell,
  Science {\bf 269},  198  (1995).

\bibitem{Bra1995.PRL75.1687}
C.~C. Bradley, C.~A. Sackett, J.~J. Tollett, and R.~G. Hulet, Phys. Rev. Lett.
  {\bf 75},  1687  (1995).

\bibitem{Dav1995.PRL75.3969}
K.~B. Davis, M.~O. Mewes, M.~R. Andrews, N.~J. van Druten, D.~S. Durfee, D.~M.
  Kurn, and W. Ketterle, Phys. Rev. Lett. {\bf 75},  3969  (1995).

\bibitem{Bra1997.PRL79.1170}
C.~C. Bradley, C.~A. Sackett, J.~J. Tollett, and R.~G. Hulet, Phys. Rev. Lett.
  {\bf 79},  1170  (1997).

\bibitem{Mat1999.PRL83.2498}
M.~R. Matthews, B.~P. Anderson, P.~C. Haljan, D.~S. Hall, C.~E. Wieman, and
  E.~A. Cornell, Phys. Rev. Lett. {\bf 83},  2498  (1999).

\bibitem{Mad2000.PRL84.806}
K.~W. Madison, F. Chevy, W. Wohlleben, and J. Dalibard, Phys. Rev. Lett. {\bf
  84},  806  (2000).

\bibitem{Mad2000.JMO47.2715}
K.~W. Madison, F. Chevy, W. Wohlleben, and J. Dalibard, J. Mod. Opt. {\bf 47},
  2715  (2000).

\bibitem{Abo2001.Sci292.476}
J.~R. Abo-Shaeer, C. Raman, J.~M. Vogels, and W. Ketterle, Science {\bf 292},
  476  (2001).

\bibitem{Ram2001.PRL87.210402}
C. Raman, J.~R. Abo-Shaeer, J.~M. Vogels, K. Xu, and W. Ketterle, Phys. Rev.
  Lett. {\bf 87},  210402  (2001).

\bibitem{Fet2009.RMP81.647}
A.~L. Fetter, Rev. Mod. Phys. {\bf 81},  647  (2009).

\bibitem{And2010.JLTP161.574}
B.~P. Anderson, J. Low Temp. Phys. {\bf 161},  574  (2010).

\bibitem{Dod1997.PRA56.587}
R.~J. Dodd, K. Burnett, M. Edwards, and C.~W. Clark, Phys. Rev. A {\bf 56},  587
  (1997).

\bibitem{Rok1997.PRL79.2164}
D.~S. Rokhsar, Phys. Rev. Lett. {\bf 79},  2164  (1997).

\bibitem{Pu1999.PRA59.1533}
H. Pu, C.~K. Law, J.~H. Eberly, and N.~P. Bigelow, Phys. Rev. A {\bf 59},  1533
   (1999).

\bibitem{Iso1999.PRA60.3313}
T. Isoshima and K. Machida, Phys. Rev. A {\bf 60},  3313  (1999).

\bibitem{Svi2000.PRL84.5919}
A.~A. Svidzinsky and A.~L. Fetter, Phys. Rev. Lett. {\bf 84},  5919  (2000).

\bibitem{Vir2001.PRL86.2704}
S.~M.~M. Virtanen, T.~P. Simula, and M.~M. Salomaa, Phys. Rev. Lett. {\bf 86},
  2704  (2001).

\bibitem{Sim2002.PRA65.033614}
T.~P. Simula, S.~M.~M. Virtanen, and M.~M. Salomaa, Phys. Rev. A {\bf 65},
  033614  (2002).

\bibitem{Kaw2004.PRA70.043610}
Y. Kawaguchi and T. Ohmi, Phys. Rev. A {\bf 70},  043610  (2004).

\bibitem{Jac2005.PRA72.053617}
A.~D. Jackson, G.~M. Kavoulakis, and E. Lundh, Phys. Rev. A {\bf 72},  053617
  (2005).

\bibitem{Huh2006.PRA74.063619}
J.~A.~M. Huhtam\"aki, M. M\"ott\"onen, and S.~M.~M. Virtanen, Phys. Rev. A {\bf
  74},  063619  (2006).

\bibitem{Lun2006.PRA74.063620}
E. Lundh and H.~M. Nilsen, Phys. Rev. A {\bf 74},  063620  (2006).

\bibitem{Cap2009.JPB42.145301}
P. Capuzzi and D.~M. Jezek, J. Phys. B: At. Mol. Opt. Phys. {\bf 42},  145301
  (2009).

\bibitem{Kuo2010.PRA81.023603}
P. Kuopanportti, E. Lundh, J.~A.~M. Huhtam\"aki, V. Pietil\"a, and M.
  M\"ott\"onen, Phys. Rev. A {\bf 81},  023603  (2010).

\bibitem{Shi2004.PRL93.160406}
Y. Shin, M. Saba, M. Vengalattore, T.~A. Pasquini, C. Sanner, A.~E. Leanhardt,
  M. Prentiss, D.~E. Pritchard, and W. Ketterle, Phys. Rev. Lett. {\bf 93},
  160406  (2004).

\bibitem{Mot2003.PRA68.023611}
M. M\"ott\"onen, T. Mizushima, T. Isoshima, M.~M. Salomaa, and K. Machida,
  Phys. Rev. A {\bf 68},  023611  (2003).

\bibitem{Gaw2006.JPhysB39.L225}
K. Gawryluk, M. Brewczyk, and K. Rz\c{a}\.zewski, J. Phys. B: At. Mol. Opt.
  Phys. {\bf 39},  L225  (2006).

\bibitem{Huh2006.PRL97.110406}
J.~A.~M. Huhtam\"aki, M. M\"ott\"onen, T. Isoshima, V. Pietil\"a, and S.~M.~M.
  Virtanen, Phys. Rev. Lett. {\bf 97},  110406  (2006).

\bibitem{Mat2006.PRL97.180409}
A. Mu\~{n}oz Mateo and V. Delgado, Phys. Rev. Lett. {\bf 97},  180409  (2006).

\bibitem{Iso2007.PRL99.200403}
T. Isoshima, M. Okano, H. Yasuda, K. Kasa, J.~A.~M. Huhtam\"{a}ki, M. Kumakura,
  and Y. Takahashi, Phys. Rev. Lett. {\bf 99},  200403  (2007).

\bibitem{Kar2009.JPhysB42.095301}
T. Karpiuk, M. Brewczyk, M. Gajda, and K. Rz\c{a}\.zewski, J. Phys. B: At. Mol.
  Opt. Phys. {\bf 42},  095301  (2009).

\bibitem{Kuo2010.PRA81.033627}
P. Kuopanportti and M. M\"ott\"onen, Phys. Rev. A {\bf 81},  033627  (2010).

\bibitem{Ino2001.PRL87.080402}
S. Inouye, S. Gupta, T. Rosenband, A.~P. Chikkatur, A. G\"{o}rlitz, T.~L.
  Gustavson, A.~E. Leanhardt, D.~E. Pritchard, and W. Ketterle, Phys. Rev.
  Lett. {\bf 87},  080402  (2001).

\bibitem{Hod2001.PRL88.010405}
E. Hodby, G. Hechenblaikner, S.~A. Hopkins, O.~M. Marag\`{o}, and C.~J. Foot,
  Phys. Rev. Lett. {\bf 88},  010405  (2001).

\bibitem{Sch2007.PRL98.110402}
D.~R. Scherer, C.~N. Weiler, T.~W. Neely, and B.~P. Anderson, Phys. Rev. Lett.
  {\bf 98},  110402  (2007).

\bibitem{Abr1995.PRB52.7018}
M. Abraham, I. Aranson, and B. Galanti, Phys. Rev. B {\bf 52},  R7018  (1995).

\bibitem{Ara1996.PRB53.75}
I. Aranson and V. Steinberg, Phys. Rev. B {\bf 53},  75  (1996).

\bibitem{Nak2000.PhysicaB284.17}
M. Nakahara, T. Isoshima, K. Machida, S.-I. Ogawa, and T. Ohmi, Physica B:
  Condens. Matter {\bf 284--288},  17  (2000).

\bibitem{Iso2000.PRA61.063610}
T. Isoshima, M. Nakahara, T. Ohmi, and K. Machida, Phys. Rev. A {\bf 61},
  063610  (2000).

\bibitem{Oga2002.PRA66.013617}
S.-I. Ogawa, M. M\"ott\"onen, M. Nakahara, T. Ohmi, and H. Shimada, Phys. Rev.
  A {\bf 66},  013617  (2002).

\bibitem{Mot2002.JPCM14.13481}
M. M\"ott\"onen, N. Matsumoto, M. Nakahara, and T. Ohmi, J. Phys.: Condens.
  Matter {\bf 14},  13481  (2002).

\bibitem{Pie2008.book.vortices}
V. Pietil\"a, M. M\"ott\"onen, and M. Nakahara,  in {\em Electromagnetic,
  Magnetostatic, and Exchange Interaction Vortices in Confined Magnetic
  Structures}, edited by E.~O. Kamenetskii (Transworld Research Network,
  Kerala, 2008).

\bibitem{Lea2002.PRL89.190403}
A.~E. Leanhardt, A. G\"{o}rlitz, A.~P. Chikkatur, D. Kielpinski, Y. Shin, D.~E.
  Pritchard, and W. Ketterle, Phys. Rev. Lett. {\bf 89},  190403  (2002).

\bibitem{Mot2007.PRL99.250406}
M. M\"ott\"onen, V. Pietil\"a, and S.~M.~M. Virtanen, Phys. Rev. Lett. {\bf
  99},  250406  (2007).

\bibitem{Xu2008.PRA78.043606}
Z.~F. Xu, P. Zhang, C. Raman, and L. You, Phys. Rev. A {\bf 78},  043606
  (2008).

\bibitem{Xu2008.NJP11.055019}
Z.~F. Xu, R.~Q. Wang, and L. You, New J. Phys. {\bf 11},  055019  (2008).

\bibitem{Xu2010.PRA81.053619}
Z.~F. Xu, P. Zhang, R. L\"u, and L. You, Phys. Rev. A {\bf 81},  053619
  (2010).

\bibitem{Kuo2010.JLTP161.561}
P. Kuopanportti and M. M\"ott\"onen, J. Low Temp. Phys. {\bf 161},  561
  (2010).

\bibitem{Pet1995.PRL74.3352}
W. Petrich, M.~H. Anderson, J.~R. Ensher, and E.~A. Cornell, Phys. Rev. Lett.
  {\bf 74},  3352  (1995).

\bibitem{Pie2007.PRA76.023610}
V. Pietil\"a, M. M\"ott\"onen, and S.~M.~M. Virtanen, Phys. Rev. A {\bf 76},  023610
  (2007).

\bibitem{Ohm1998.JPSJ67.1822}
T. Ohmi and K. Machida, J. Phys. Soc. Jpn. {\bf 67},  1822  (1998).

\bibitem{Ho1998.PRL81.742}
T.-L. Ho, Phys. Rev. Lett. {\bf 81},  742  (1998).

\bibitem{Ued2010.book.Bose}
M. Ueda, {\em Fundamentals and New Frontiers of Bose-Einstein Condensation}
  (World Scientific, Singapore, 2010).

\bibitem{Hod2000.JPhysB33.4087}
E. Hodby, G. Hechenblaikner, O.~M. Marag\'o, J. Arlt, S. Hopkins, and C.~J.
  Foot, J. Phys. B: At. Mol. Opt. Phys. {\bf 33},  4087  (2000).

\bibitem{Hod2001.PRL86.2196}
E. Hodby, O.~M. Marag\`o, G. Hechenblaikner, and C.~J. Foot, Phys. Rev. Lett.
  {\bf 86},  2196  (2001).

\bibitem{Lea2003.PRL90.140403}
A.~E. Leanhardt, Y. Shin, D. Kielpinski, D.~E. Pritchard, and W. Ketterle,
  Phys. Rev. Lett. {\bf 90},  140403  (2003).

\bibitem{Kum2006.PRA73.063605}
M. Kumakura, T. Hirotani, M. Okano, Y. Takahashi, and T. Yabuzaki, Phys. Rev. A
  {\bf 73},  063605  (2006).

\bibitem{Oka2007.JLTP148.447}
M. Okano, H. Yasuda, K. Kasa, M. Kumakura, and Y. Takahashi, J. Low Temp. Phys.
  {\bf 148},  447  (2007).

\bibitem{Shi2011.JPB44.075302}
H. Shibayama, Y. Yasaku, and T. Kuwamoto, J. Phys. B: At. Mol. Opt. Phys. {\bf
  44},  075302  (2011).

\bibitem{Sim2005.PRL94.080404}
T.~P. Simula, P. Engels, I. Coddington, V. Schweikhard, E.~A. Cornell, and
  R.~J. Ballagh, Phys. Rev. Lett. {\bf 94},  080404  (2005).

\bibitem{Nee2010.PRL104.160401}
T.~W. Neely, E.~C. Samson, A.~S. Bradley, M.~J. Davis, and B.~P. Anderson,
  Phys. Rev. Lett. {\bf 104},  160401  (2010).

\bibitem{Kla2001.PRA64.053602}
N.~N. Klausen, J.~L. Bohn, and C.~H. Greene, Phys. Rev. A {\bf 64},  053602
  (2001).

\bibitem{Kem2002.PRL88.093201}
E.~G.~M. van Kempen, S.~J. J. M.~F. Kokkelmans, D.~J. Heinzen, and B.~J.
  Verhaar, Phys. Rev. Lett. {\bf 88},  093201  (2002).

\bibitem{Wid2006.NJP8.152}
A. Widera, F. Gerbier, S. F\"olling, T. Gericke, O. Mandel, and I. Bloch, N. J.
  Phys {\bf 8},  152  (2006).

\bibitem{Note1}
The atom--atom interactions are not particularly important for the
  operation of the pump, and qualitatively similar results are obtained, e.g.,
  for weak enough antiferromagnetic coupling ($g_\protect \mathrm {s} > 0$) and in the
  noninteracting case.

\bibitem{Note2}
The parameter values in the simulations have been chosen such that the 
ratio of magnetic to interaction energy is of the same order as in the experimental setup 
outlined in Sec.~\ref{sc:experimental}.


\bibitem{Ber1984.ProcRSocA392.45}
M.~V. Berry, Proc. R. Soc. A {\bf 392},  45  (1984).

\end{thebibliography}
\end{document}